\documentclass{PoS}

\usepackage{enumerate}
\usepackage{amsmath,amssymb}  
\usepackage{bm}               
\usepackage{amscd}            
\usepackage{graphicx}         
\usepackage{epsfig}
\usepackage{hhline,multirow}  
\usepackage{dcolumn}          
\usepackage[dvipsnames]{xcolor}
\usepackage[normalem]{ulem}
\usepackage{mathtools}

\usepackage{slashed}

\newcommand{\nbar}{{\overline n}}
\newcommand{\nslash}{n\hspace*{-0.22cm}\slash\hspace*{0.022cm}}
\newcommand{\nbslash}{\nbar\hspace*{-0.22cm}\slash\hspace*{0.022cm}}
\newcommand{\Tr}{{\rm Tr}}

\newcommand{\vpsq}{v^{\,\prime\,2}}

\title{Hadron Mass Effects on Kaon Multiplicities: HERMES vs. COMPASS}

\ShortTitle{Hadron Mass Effects: Kaons at HERMES vs. COMPASS}

\author{\speaker{Juan V. Guerrero} and Alberto Accardi \\
	Hampton U. and Jefferson Lab, USA\\
	E-mail: \email{juanvg@jlab.org}}

\abstract{
Experimental data for integrated kaon multiplicities taken at HERMES and COMPASS measurements look incompatible with each other. In this talk, we investigate the effects of hadron masses calculated at leading-order and leading twist at the kinematics of these two experiments. We present evidence that Hadron Mass Corrections can fully reconcile the data for the $K^+/K^-$ multiplicity ratio, and can also sizeably reduce the apparent large discrepancy in the case of $K^+ + K^-$ data. Residual differences in the shape of the latter one remains to be understood.
}

\FullConference{QCD Evolution 2017\\
	22-26 May, 2017\\
	Jefferson Lab, Newport News, VA - USA}

\begin{document}
	
\section{Introduction}

The strange quark Parton Distribution Function (PDF) is very important because it appears in any calculation involving light-quarks. Currently, there exist several sets of PDFs with small uncertainties for the valence quarks and gluons. In contrast to this, the strange PDF has been 
experimentally investigated by several collaborations like HERMES at HERA \cite{Airapetian:2008qf,JR:2013qra,Airapetian:2013zaw}, ATLAS and CMS at the LHC \cite{Aad:2012sb,Chatrchyan:2013mza}, or determined in global PDF fits by several groups \cite{Ball:2014uwa,Harland-Lang:2014zoa,Dulat:2015mca}, all of these show large discrepancies in size and shape.

There are several ways to access the strange quark PDFs, in particular by analyzing
 $z-$integrated multiplicities in Semi-Inclusive Deeply Inelastic Scattering (SIDIS) on deuteron targets. These multiplicities have been measured by the  HERMES \cite{Airapetian:2013zaw,Airapetian:2012ki} and COMPASS \cite{Seder:2015sdw,Adolph:2016bwc} collaborations showing large discrepancies between their measurements. However, these measurements are sensitive to relatively low values of photon virtualities ($Q \approx 1-4$ GeV) where the mass $m$ of the target nucleon and the mass $m_h$ of the observed hadron, in this case the Kaon ($m_K\approx 0.5$ GeV), induce non-negligible 
``Hadron Mass Corrections'' (HMCs) of order ${\mathcal{O}(m^2/Q^2)}$ \cite{Accardi:2009md,Guerrero:2015wha}.

In this talk, we will show results which quantify these HMCs for Kaon multiplicities in electron deuteron Semi-inclusive Deep Inelastic Scattering (SIDIS) at HERMES and COMPASS. We present evidence that these are not negligible, and may be largely responsible for the apparent discrepancies between the measurements performed by the two collaborations. 
	
\section{Leading order multiplicities at finite $Q^2$}

\begin{figure}[bt]
  \centering
  \parbox[c]{5.5cm}{\includegraphics[width=\linewidth]{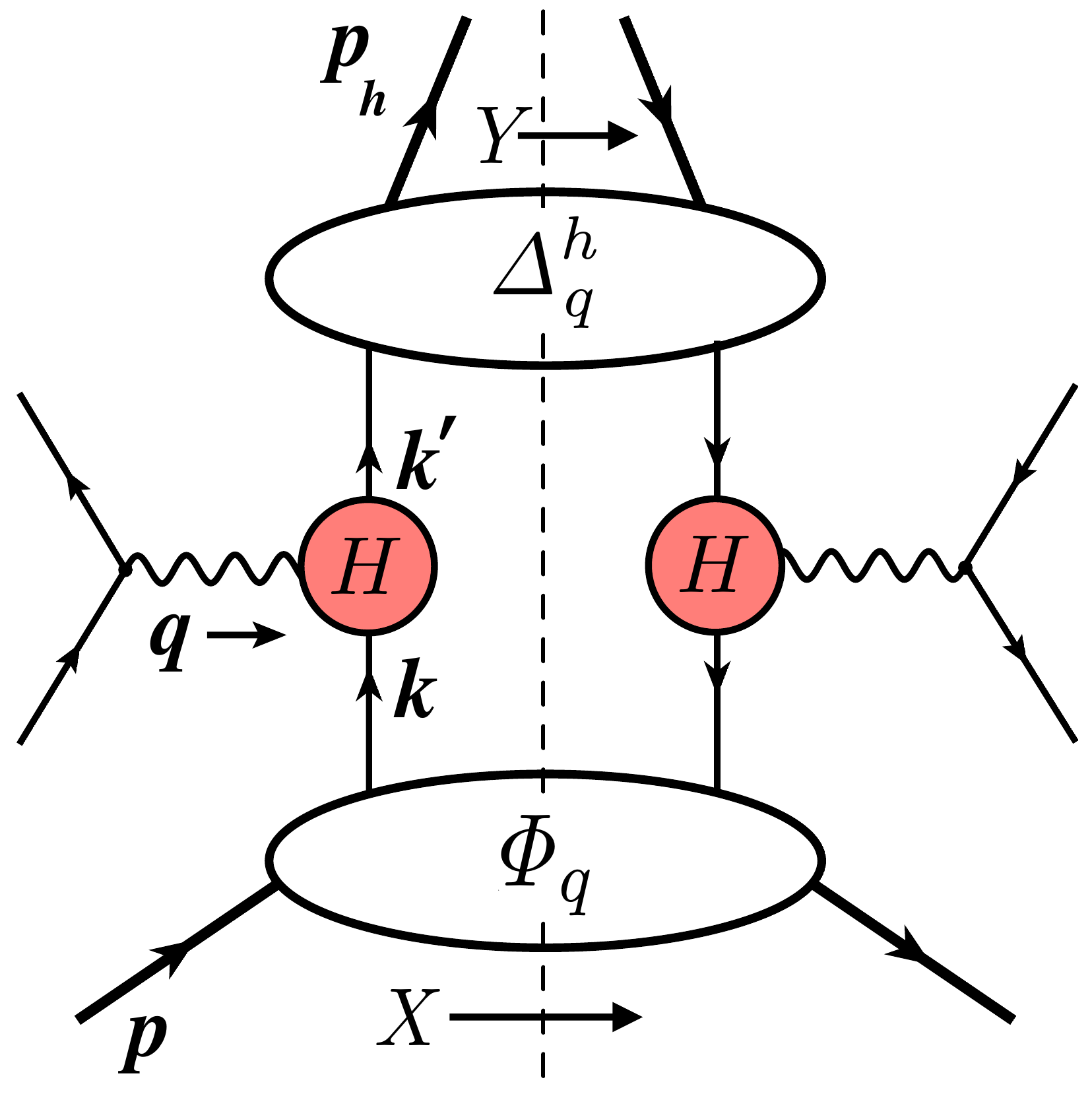}}
  \hspace{1cm}
  \parbox[c]{5.5cm}{\includegraphics[width=\linewidth]{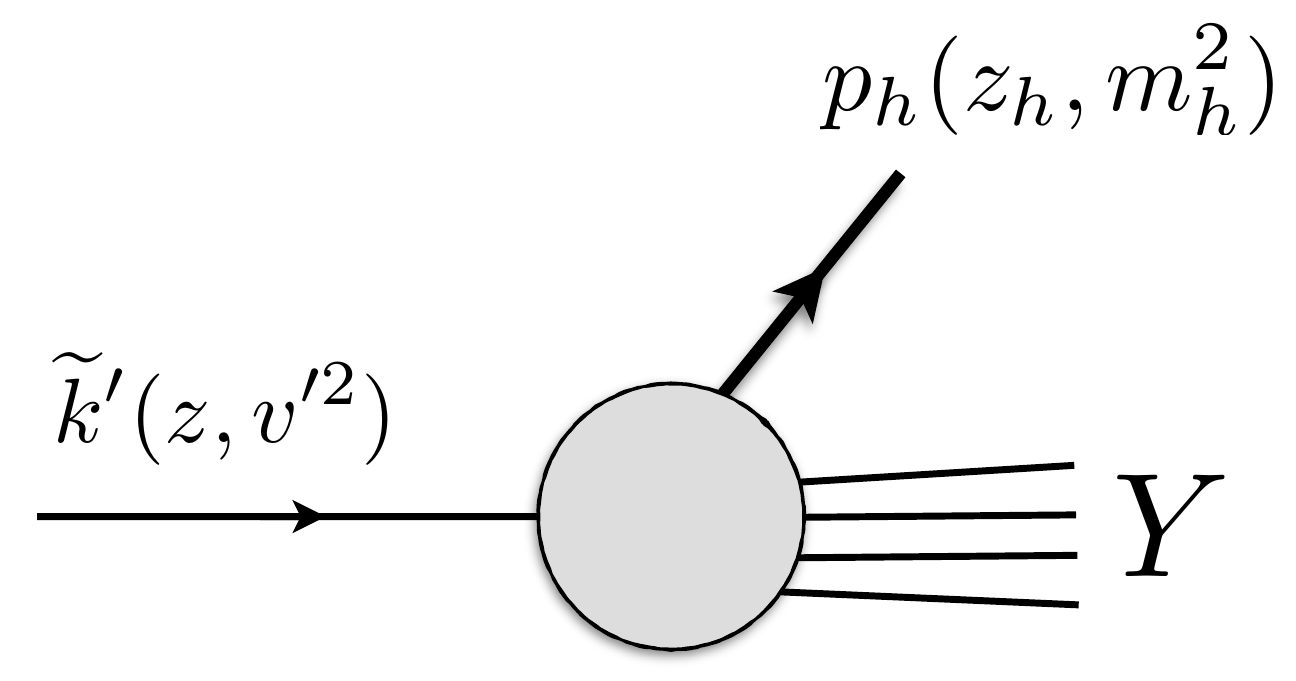}\\[.5cm]}     
  \caption{{\it Left:} SIDIS handbag diagram and kinematics, where $q$ is the momentum of the photon, $p$ of the target nucleon, $p_h$ of the observed hadron, $k$ and $k'$ of the partons participating in the hard scattering $H$. {\it Right:} fragmentation vertex with factorized kinematics.}
  \label{fig:SIDIS_handbag}
\end{figure}

The $z$-integrated hadron multiplicities measured by the HERMES and COMPASS collaborations are defined as a ratio of the semi-inclusive to inclusive cross sections,
\begin{equation}
M^{h}(x_B^{\textrm exp}) = \frac{\int_{\textrm exp} dx_B dQ^2 \int_{0.2}^{0.8(0.85)}\, dz_h\, \frac{d\sigma^h}{dx_B dQ^2 dz_h}}{\int_{\textrm exp} dx_B dQ^2  \frac{d\sigma^{\textrm{DIS}}}{dx_B dQ^2}} \ ,
\label{eq:def_multiplicities}
\end{equation}
where $x_B= \frac{Q^2}{2 p \cdot q}$ and $Q^2 = -q^2$, namely the Bjorken scaling variable
and the virtuality of the exchanged photon respectively, are the usual inclusive invariants,
$z_h= \frac{p \cdot p_h}{p \cdot q}$ is the fragmentation invariant, and the rest of kinematics variables are defined in Fig.~\ref{fig:SIDIS_handbag}
left\footnote{In fact, COMPASS defines integrated multiplicities as averages over $y$ of the differential ones $\int dz_h \langle M^h(x_b,y,z_h)\rangle_y$ without precisely defining the average symbol; in this talk we will use Eq.~\eqref{eq:def_multiplicities} for both experiments.}. The integration over the inclusive invariants, $dx_B dQ^2$, is performed over the bin of nominal value $x_B^{\textrm exp}$ with the integration over $dQ^2$ being performed within $x_B$-dependent limits defined by each experiment's kinematic cuts~\cite{Seder:2015sdw,Aschenauer:2015rna}; more details are discussed in Ref.~\cite{Guerrero:2017yvf}. The $z_h-$integration limits are those defined for each experiment, for which we denote COMPASS with a parenthesis. 

In collinear factorization, only the light-cone components of momenta enter the cross sections. Therefore, we consider massive scaling variables defined by the relevant light-cone fractions $\xi= q+/p+$, also known as Nachtmann scaling variable, and $\zeta_h = p_h^-/q^-$. In the so-called $(p,q)$ frame, where  $p$ and $q$ are collinear and have zero transverse momentum, one finds \cite{Guerrero:2015wha}
\begin{align}
\label{eq:xi}
\xi & \equiv -\frac{q^+}{p^+}
= \frac{2 x_B}{1 + \sqrt{1 + 4 x_B^2 M^2/Q^2}} \\
\label{eq:zeta}
\zeta_h & \equiv \frac{p_h^-}{q^-} = \frac{z_h}{2} \frac{\xi}{x_B}
\left( 1 + \sqrt{1 - \frac{4 {x_B^2} M^2 m_{h}^2}{{z_h^2}\ Q^4}} 
\right) \ ,
\end{align}
where $M$ is the nucleon target mass and $m_h$ is the detected hadron mass. 
Note that in the Bjorken limit,where $M^2/Q^2 \to 0$ and $m_h^2/Q^2 \to 0$, we recover the usual massless scaling variables $x_B$ and $z_h$ .

The hadronic tensor for SIDIS at Leading order (LO) in the strong coupling constant, see Fig.~\ref{fig:SIDIS_handbag}, can be expressed in terms of quark-quark correlators, $\Phi_q$ and $\Delta_q^h$, related with the quark distribution and fragmentation functions, respectively \cite{Collins:1989gx, Mulders01}, and reads
\begin{equation}
W^{\mu\nu} \propto 
\sum_q e_q^2 \int d^4k\ d^4k'\
\Tr \left[ \Phi_q(p,k)\, \gamma^\mu\, \Delta_q^h(k',p_h)\, \gamma^\nu
\right] \, \delta^{(4)}(k + q - k') \, .
\label{eq:Wmunu}
\end{equation}

Getting a factorized expression for the hadronic tensor Eq.~\eqref{eq:Wmunu} requires two steps.
First, the quark-quark correlators need to be expanded in inverse powers of $k^+$ and $k'^-$, 
which are the leading components of the parton momenta incoming and outcoming the hard scattering $H$ in Figure~\ref{fig:SIDIS_handbag}. Keeping only the first order of this expansion, namely the ``leading-twist (LT)'' terms, the expanded correlators looks like
$\Phi = k^+ \big[ \phi_2(k) \nbslash + {\cal O}(1/k^+) \big]$
and 
$\Delta = k'^+ \big[ \delta_2(k') \nslash + {\cal O}(1/k'^-) \big]$. Here, $ \phi_2 $ and $\delta_2$ are scalar functions of the momenta, and $n^\mu$ and $\bar n^\mu$ are the unit light-cone plus- and minus-vectors, respectively. Then, 
\begin{align}
	W^{\mu\nu} \propto \int d^4k\ d^4k'\ 
	\Tr \left[ \nbslash \, \gamma^\mu\, \nslash \,
	\gamma^\nu \right] \, \phi_2(k) \delta_2(k') k^+ k'^- \delta^{(4)}( k + q -  k') + \,{\rm HT} \ , 
	\label{eq:Wmunu-2}
\end{align}
with HT standing for Higher-Twist (HT) contributions\footnote{The Higher-Twist contributions neglected here are not forgotten, actually they contribute to restore gauge invariance in HT diagrams that include the exchange of an extra parton between $H$ and $\phi_q$ or $\Delta_q^h$ in Fig.~\ref{fig:SIDIS_handbag} \cite{Bacchetta:2006tn}.}. The hadronic tensor in Eq.~\eqref{eq:Wmunu-2} satisfies the Ward identity $q_\mu W^{\mu \nu}=0$, meaning that at LO and at LT our scheme is gauge invariant. A more detailed discussion of our scheme and its relation with the parton model can be found in Ref.~\cite{Guerrero:2017yvf}. 

Now, we go to the second step which consist in making a collinear approximation to the momenta of the scattered and fragmenting parton namely, $k \approx \widetilde{k}$ and  $k'\approx \widetilde{k'}$, where  $\widetilde{k}$ is collinear to the target nucleon and $\widetilde{k'}$ is collinear to the detected hadron. This approximation is made only in the $\delta-$function in Eq.~\eqref{eq:Wmunu-2}, $\delta^{(4)}( k + q - k') \to \delta^{(4)}(\widetilde k + q - \widetilde k') $. We parametrize the approximated four-momenta of the initial and scattered parton as
\begin{align}
k^{\mu} \approx \widetilde k^{\mu}
& = \Big(xp^+,\frac{v\,^2}{2xp^+},\bm{0}_T\Big) \label{eq:k} \\
k'^{\mu} \approx \widetilde k'^{\mu}
& = \Bigg(\frac{\vpsq +
	(\bm{p_{h\perp}}/z)^2}{2p_h^-/z},
\frac{p_h^-}{z},\frac{\bm{p_{h\perp}}}{z}\Bigg) \ .
\label{eq:kp}    
\end{align}
where $x = \frac{k^+}{p^+}$ and $z = \frac{p_h^-}{k'^-}$, and the ``average virtualities'' $v\,^2 \approx \langle k_\mu k^\mu\rangle$ and $\vpsq \approx \langle k'_\mu k^{\prime \mu}\rangle$  will be fixed later. An important remark is that these virtualities are determined by the dynamics of the scattering and hadronization process without the need, in principle, to be equal to the current mass of the quarks. Integrating over the hard scattering vertex, the $\delta-$function imposes four momentum approximation for the approximate quark momenta and sets
\begin{subequations}
	\begin{eqnarray}
	\frac{x}{\xi} & = & 1 + \frac{z}{\zeta_h}\frac{v\,'\,^2}{Q^2} \label{eq:hard_scattering_I} \\
	\frac{\zeta_h}{z} & = & 1 + \frac{\xi}{x}\frac{v\,^2}{Q^2} \ .	\label{eq:hard_scattering_II}
	\end{eqnarray}
	\label{eq:Hard_scattering}
\end{subequations}

Now, we need to discuss the choice of virtualities $v\,^2 $ and $v\,'\,^2$. Assuming as usual that $v\,^2 = 0$ (which is a kinematically allowed value of $k_\mu k^\mu$, see Ref.~\cite{Guerrero:2015wha,Guerrero:2017yvf}) one obtains $z=\zeta_h$ and $x= \xi \Big( 1 + \frac{v\,'\,^2}{Q^2} \Big).$ On the other hand, the scattered parton is fragmenting into a massive hadron, and therefore needs a non vanishing virtuality $v\,'\,^2$. 
In order to get the minimal virtuality required we match the partonic kinematic with hadronic kinematics in the fragmentation process. This constrains the virtuality to be at least $v\,'\,^2=m_h^2/\zeta_h$. Finally, we obtain
\begin{eqnarray}
  x &=& \xi_h \equiv \xi \Big(1 + \frac{m_h^2}{\zeta_h Q^2} \Big)
        \label{eq:xi_h} \\
  z &=& \zeta_h   \ .
        \label{eq:zeta_h}
\end{eqnarray} 
where the Nachtmann-type scaling variables $\xi$ and $\zeta_h$  were defined previously in Eqs.~\eqref{eq:xi}-\eqref{eq:zeta}.

Then, the LO finite-$Q^2$ $z$-integrated hadron multiplicity can be written as a factorized expression in terms of quark PDFs, $q$, and FFs, $D_q^h$, evaluated at the scaling variables $\xi_h$ and $\zeta_h$ just derived:
\begin{equation}
  M^{h}(x_B^{\textrm exp}) = \frac{\sum_q e_q^2\, \int_{\textrm exp} dx_B dQ^2 \int_{0.2}^{0.8(0.85)}\, dz_h\, J_h \,  q(\xi_h,Q^2)\, D_q^h(\zeta_h,Q^2)}{\sum_q e_q^2\, \int_{\textrm exp} dx_B dQ^2 \, q(\xi,Q^2)} \ ,
  \label{eq:finite_Q2_multiplicities}
\end{equation}
where $J_h$ is a Jacobian factor \cite{Guerrero:2015wha}. Note that in the Bjorken limit, Eq.~(\ref{eq:finite_Q2_multiplicities}) reduces to the usual, ``massless'' $M_h^{(0)}$ multiplicity,
\begin{equation}
  M^{h (0)}(x_B^{\textrm exp}) = \frac{\sum_q e_q^2\, \int_{\textrm exp} dx_B dQ^2 \, q(x_B, Q^2)\,\int_{0.2}^{0.8(0.85)} dz_h D_q^h(z_h, Q^2)}{\sum_q e_q^2 \int_{\textrm exp} dx\,dQ^2\,  q(x_B, Q^2)} \ .
  \label{eq:partonlevel_multiplicities}
\end{equation}

\section{Integrated kaon multiplicities}

The HERMES and COMPASS measurements \cite{Airapetian:2012ki,Seder:2015sdw,Adolph:2016bwc} for integrated kaon multiplicities do not appear to be compatible with each other, a well known fact, but discussed mainly focusing 
on  kinematic and binning issues \cite{Aschenauer:2015rna,Stolarski:2014jka,Leader:2015hna}.
In this section we discuss how this discrepancy may be in fact apparent and largely due to mass effects. These play an essential role due to the relative low $Q^2$ values dominating the HERMES and COMPASS $x_B$ bins.

One way to compare HERMES multiplicities to COMPASS multiplicities is by using the ratio between experimental data and theory prediction, because the differences in kinematic cuts and $Q^2$ evolution between the two experiments mainly cancel. We calculated and plotted these in Fig. 3 of Ref.~\cite{Guerrero:2017yvf}, using different LO sets of PDFs (MSTW08, CJ15, CT14) \cite{Martin:2009iq,Accardi:2016qay,Dulat:2015mca} and FFs (DSS07, HKNS07) \cite{deFlorian:2007aj,Hirai:2007cx}. There, one can observe a large FF systematic uncertainty, which is due to the poor knowledge we currently have of kaon fragmentation functions. After considering HMCs, the data over theory ratios become flatter, in particular for the COMPASS data. 

\subsection{Multiplicities in a massless world}

In order to make a data-to-data comparison of HERMES and COMPASS results, we define ``theoretical correction ratios''. These make the data from different experimental beam energies
directly comparable by producing approximate massless parton multiplicities at a common beam energy. They also reduce the theoretical systematic uncertainties (PDFs and FFs choice), and allow one to interpret the corrected multiplicities as parton model multiplicities using Eq.~(\ref{eq:partonlevel_multiplicities}).

This method consists of two steps. First, we remove the mass effects from the original data multiplying it by the ``HMC ratio'',
\begin{equation}
  R^h_{HMC} = \frac{M^{h(0)}}{M^h} \ ,
  \label{eq:R_HMC}
\end{equation}
where the numerator is the massless hadron multiplicity, $M^{h(0)}$, defined theoretically by Eq.~(\ref{eq:partonlevel_multiplicities}) and the denominator is 
the finite-$Q^2$ multiplicity, $M^{h}$, defined by Eq.~(\ref{eq:finite_Q2_multiplicities}). 
Then, we can interpret the product $M_{exp}^h\times R^h_{HMC}$ as a ``massless'' experimental multiplicity. In other words, in a world where nucleons and kaons were massless, this is the multiplicity that one would expect to measure.

The second step of this method consists in addressing the difference in the $Q^2$ reach of each $x_B$ bin of HERMES and COMPASS, often referred as ``evolution effects".  In this case, we choose to compare the data at COMPASS kinematics. Then, we define an evolution ratio $R^{H \rightarrow C}_{evo}$ that ``brings'' HERMES data to COMPASS energy,
\begin{equation}
  R^{H \rightarrow C}_{evo} = \frac{M^{h(0)}(x_B^{HERMES})\Bigr\rvert_{\textrm{COMPASS cuts}}}
     {M^{h(0)}(x_B^{HERMES})\Bigr\rvert_{\textrm{HERMES cuts}}} \ .
  \label{eq:R_evo}
\end{equation}

Here the numerator is the massless multiplicity calculated at a fixed $x_B$ bin in HERMES, but using the experimental kinematic cuts in $(x_B, Q^2)$ of the COMPASS experiment, and the denominator is the massless multiplicity integrated using the original HERMES kinematic cuts (see Figure 2 of Ref.~\cite{Guerrero:2017yvf} and the related discussion therein). After removing the mass effects from both sets of data using Eq.~\eqref{eq:R_HMC} and multiplying the massless HERMES multiplicity by this evolution ratio, we can define the massless and evolved (at COMPASS $Q^2$) multiplicities as,
\begin{subequations}
  \begin{eqnarray}	
    M_{exp}^{h(0)} & \equiv \ M_{exp}^h \times R^h_{HMC}\hspace{2.07cm}   &\textrm{(for COMPASS)}
    \label{eq:COMPASS_corrected} \\ 
    M_{exp}^{h(0)} & \equiv \ M_{exp}^h \times R^h_{HMC} \times R_{evo}^{H \rightarrow C} \hspace{0.74cm}  & \textrm{(for HERMES)}.
      \label{eq:HERMES_corrected}
  \end{eqnarray}
  \label{eq:data_sum_corrected}%
\end{subequations}

\begin{figure}
	\centering
	\includegraphics[width=8cm]{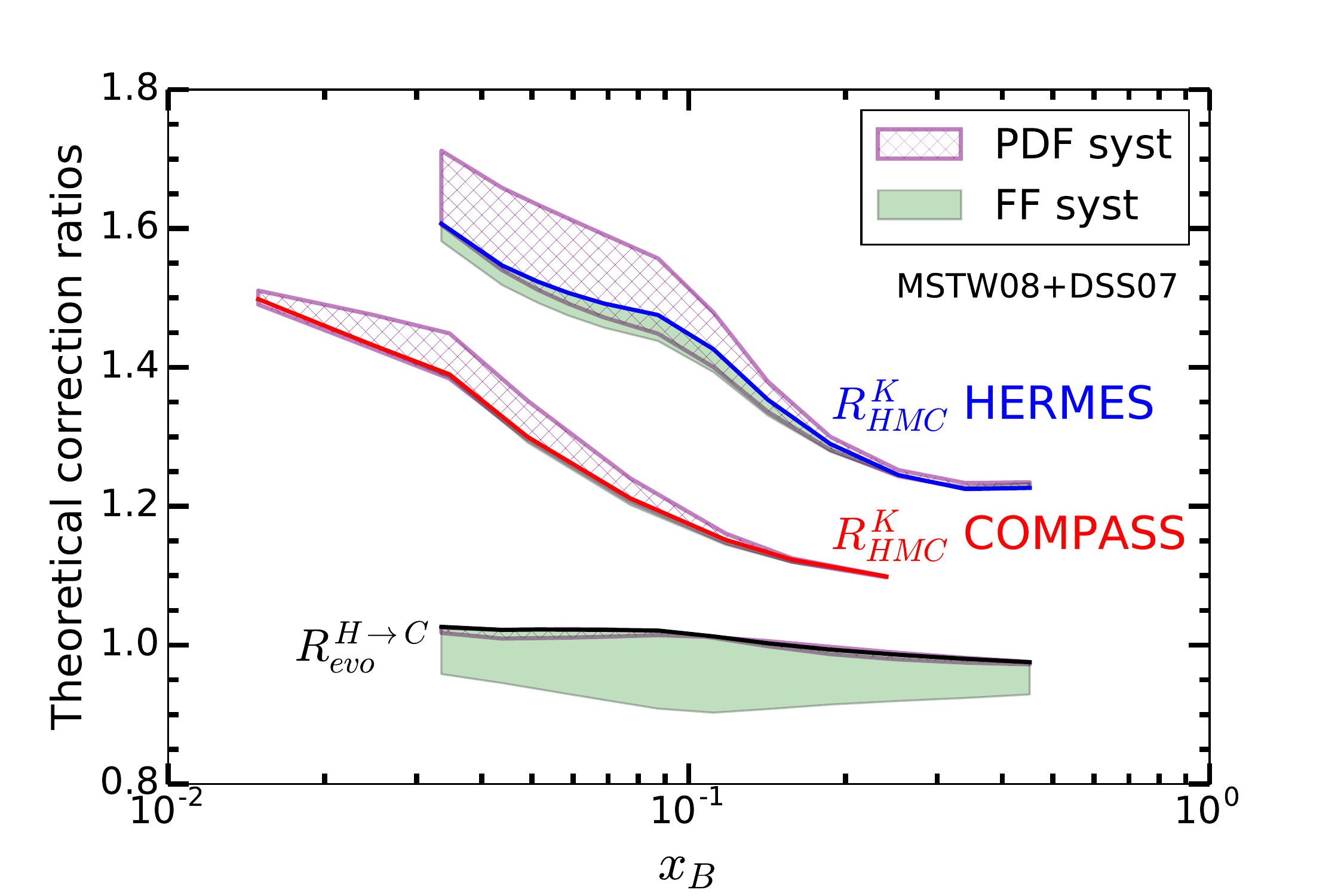}		
	\caption{Theoretical correction ratios as a function of $x_B$ for charged $K^+ + K^-$ multiplicity. The red line correspond to the mass corrections for COMPASS, the blue line correspond to the mass corrections for HERMES while the black line is the HERMES to COMPASS evolution. The green FF systematic uncertainty band for the COMPASS $R^{K}_{HMC}$ is very small compared to the HERMES case and almost invisible in the plot. The purple hashed PDF systematic uncertainty band for $R_{evo}^{H \rightarrow C}$ is very small compared to the FF systematic uncertainty.}
	\label{correction_ratios}
\end{figure}

The correction ratios are plotted in Fig.~\ref{correction_ratios}, where we find that hadron mass effects are dominant compared to evolution effects. For COMPASS, the corrections are smaller than at HERMES because the $Q^2$ accessed at COMPASS is higher bin by bin than at HERMES due to the higher beam energy. 

\begin{figure}
  \centering
  \includegraphics[width=7cm]{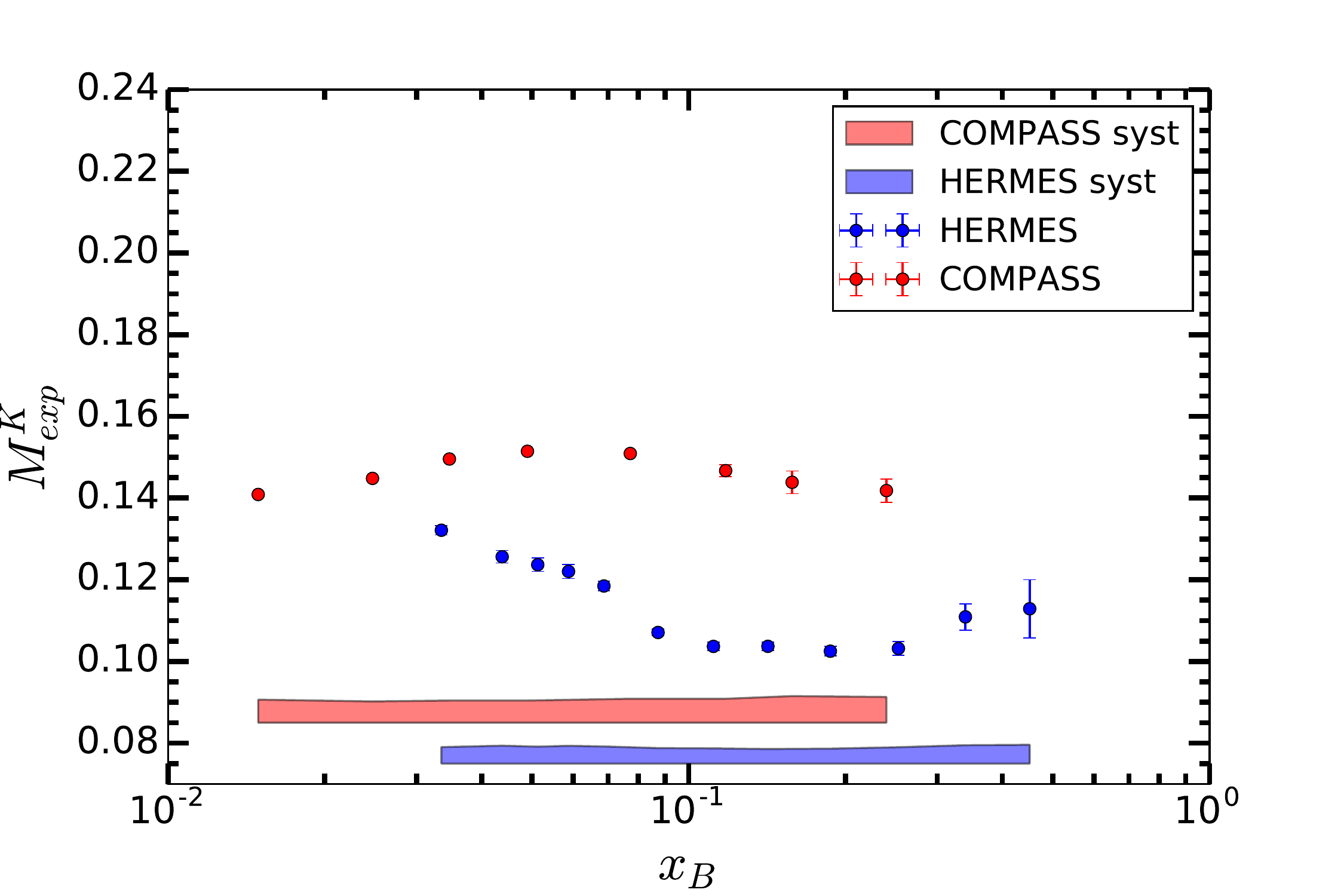}	
  \includegraphics[width=7cm]{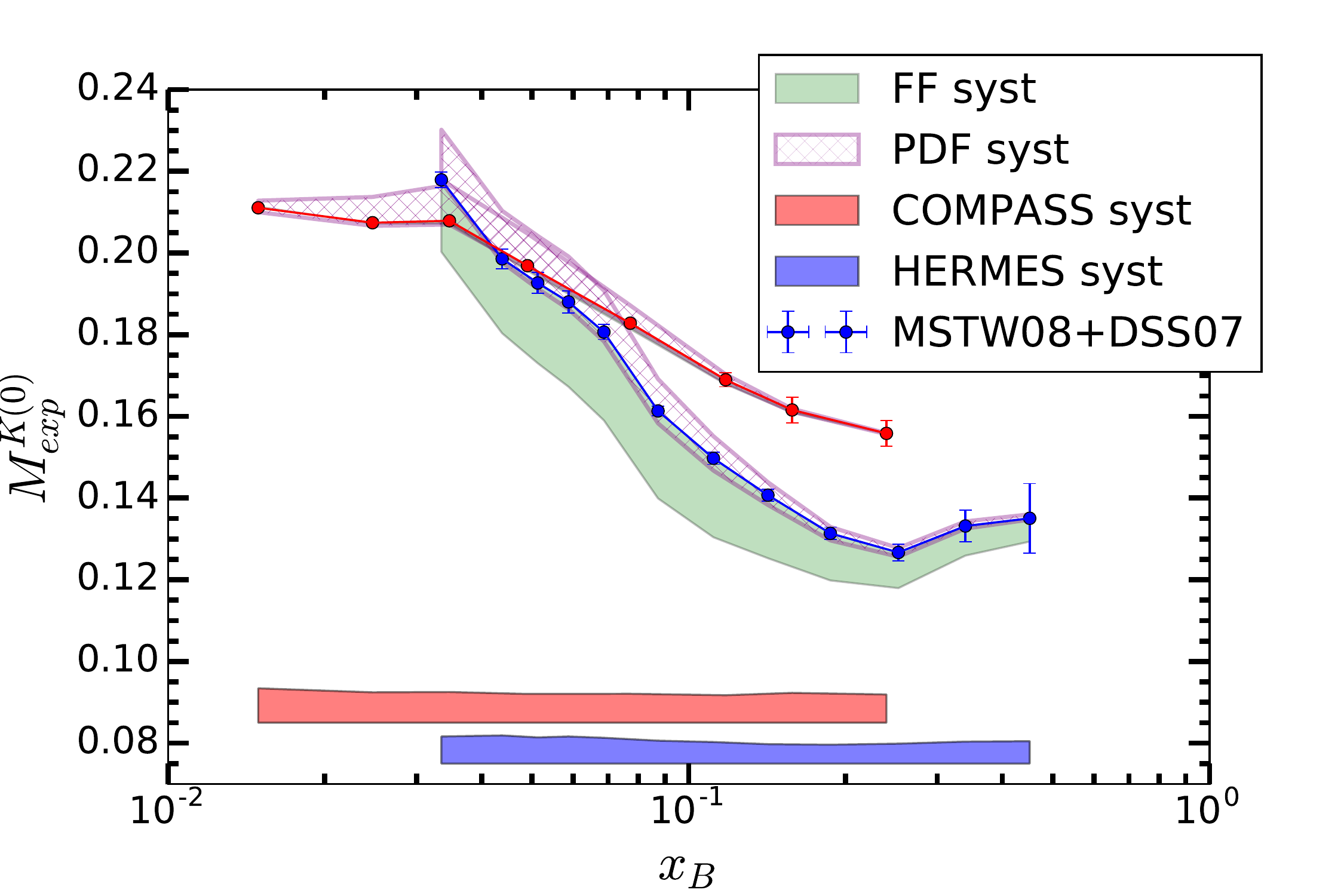}		
  \caption{Right: Experimental data for integrated kaon Multiplicities ($K^+ + K^-$). Left: Massless multiplicities at a common $Q^2$ after applying the theoretical correction ratios given by Eq.~(\ref{eq:data_sum_corrected}) to the data shown on the right.} 
  \label{corrected_data_sum}
\end{figure}

In Fig.~\ref{corrected_data_sum}, we plot the experimental $K^+ + K^-$ multiplicity data $M_{exp}^K$ on the left and the ``massless'' multiplicities $M_{exp}^{K (0)}$ on the right using Eqs.~(\ref{eq:COMPASS_corrected})-(\ref{eq:HERMES_corrected}). An important remark is that corrections are relatively stable with respect to FF and PDF choice, because the related systematic uncertainties are canceled in the correction ratios defined by Eqs.~(\ref{eq:R_HMC}) and ~(\ref{eq:R_evo}). In a way, the right plot shows how the experimental data would look like in a massless world, where the kaon multiplicities can be directly interpreted in terms of the parton model framework. The corrected data also shows a show a negative slope in $x_B$ that agrees much better with the $(1-x)^\beta$ power law behavior of any PDF, including the s-quark. There are still some discrepancy in the $x_B$ slopes and shapes of the two experimental measurements. This indicates that additional effects may play a role on top of the HMCs, or that undetected systematic uncertainties are affecting the measurements. 

\begin{figure}
	\centering	
	\includegraphics[width=7cm]{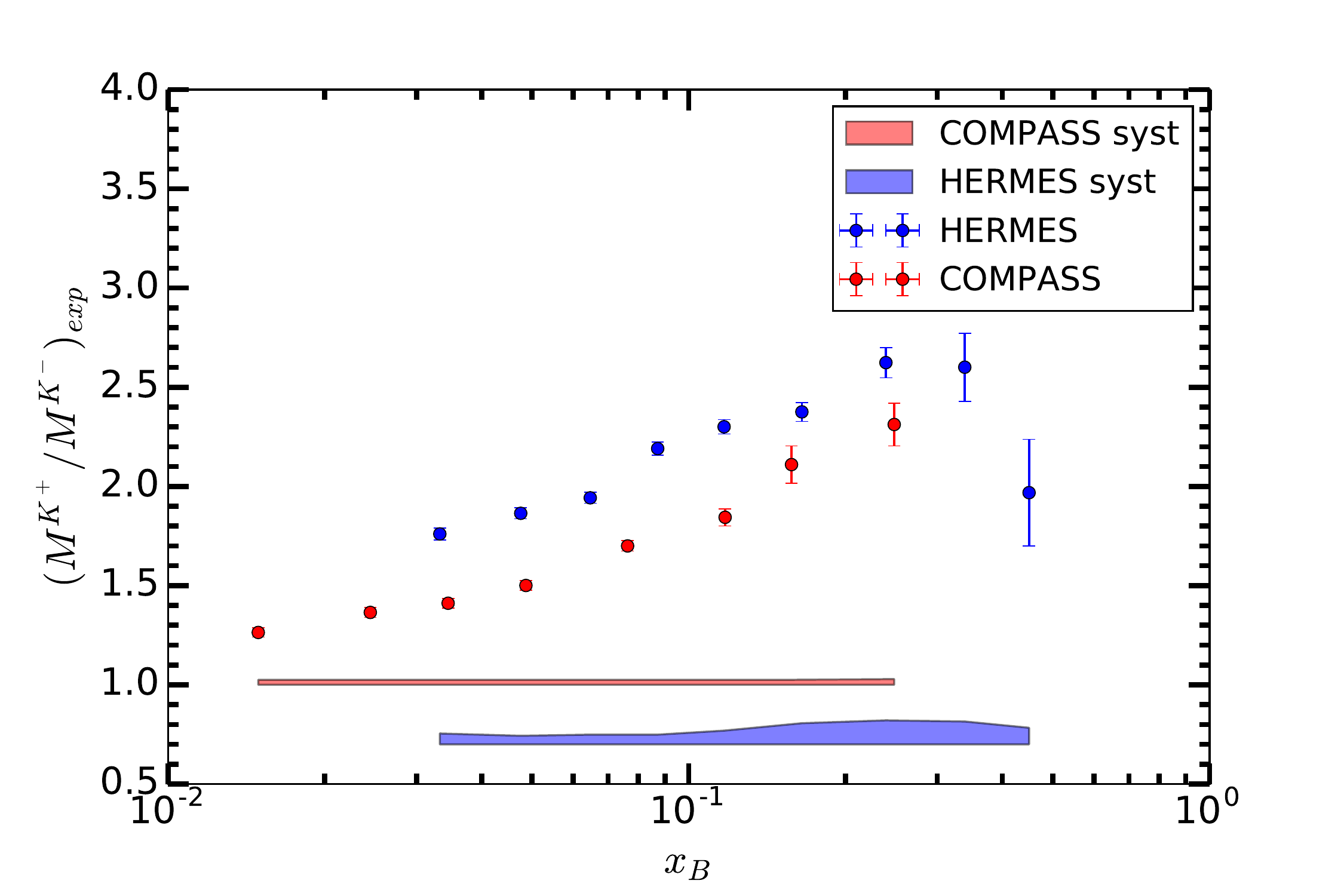}	
	\includegraphics[width=7cm]{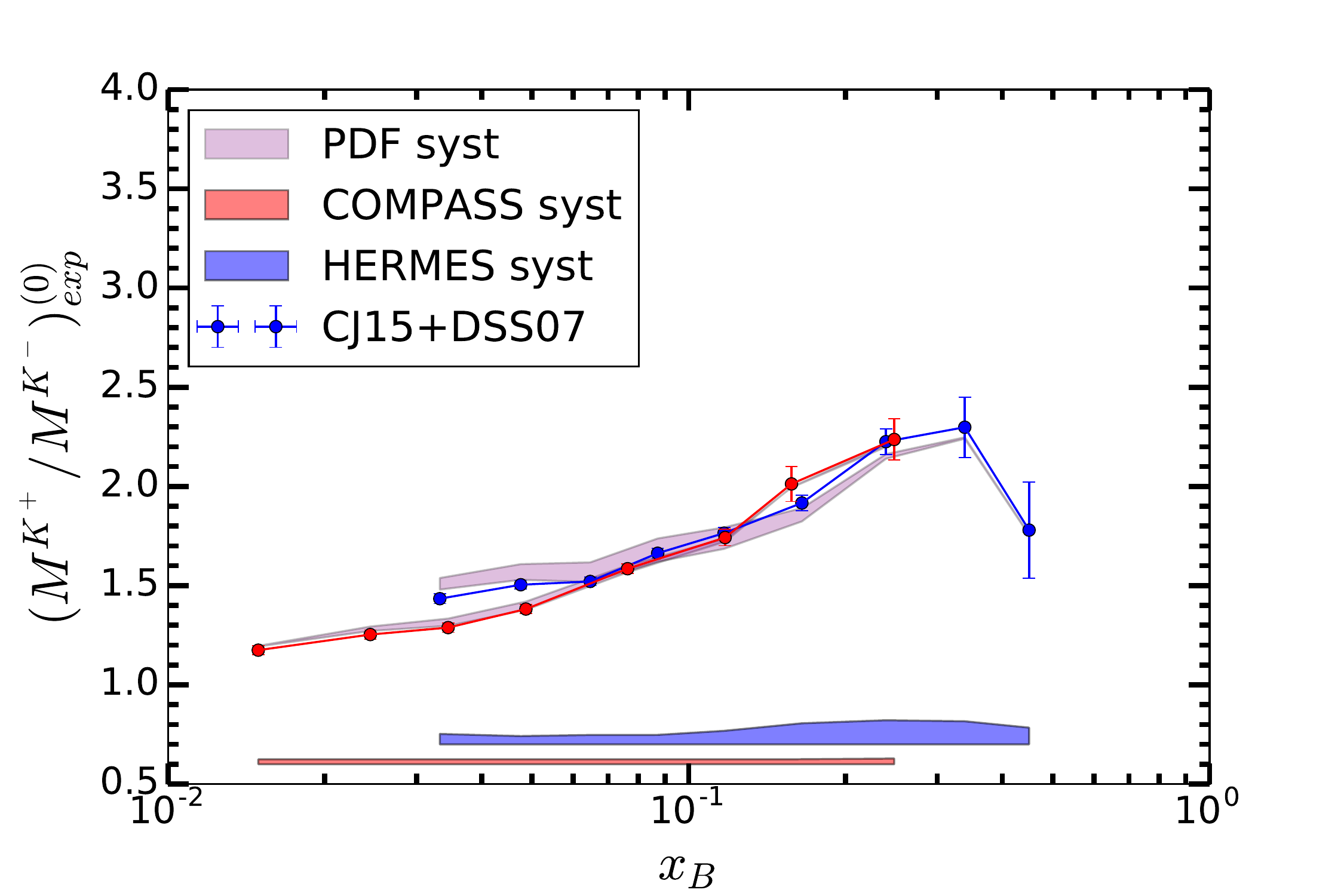}	
	\caption{Right: Experimental data for integrated kaon Multiplicities ($K^+ / K^-$). Left: Massless multiplicities at a common $Q^2$ after applying the theoretical correction ratios given by Eq.~(\ref{eq:data_sum_corrected}) to the data shown on the right.}
	\label{corrected_data_ratio}
\end{figure}

\subsection{Kaon multiplicity ratios}

Another interesting observable which can be studied is the $K^+/K^-$ multiplicity ratio. In this case the experimental systematic uncertainties  and evolution effects are expected to largely cancel in the ratio, as well as theoretical uncertainties like next-to-leading order or higher-twist effects. However, for Kaons there still are residual HMCs, although smaller than for the $K^+ + K^-$ sum, due to the difference in fragmentation functions between $K^+$ and $K^-$ (see Fig. 6 of Ref.~\cite{Guerrero:2017yvf}).

The original and ``massless'' $K^+/K^-$ data, for both HERMES and COMPASS experiments, are plotted in the left and right panels of Fig.~\ref{corrected_data_ratio}. In this case, the slopes are already compatible in the original data but there is a discrepancy in size. After removing the mass effects, the ``massless'' experimental kaon ratios become fully compatible between the two experiments except maybe last HERMES $x_B$ bin, which seems to have a drastic change in slope, as it was the case also for the sum $K^+ + K^-$. Unfortunately, this ``hockey stick'' shape lies just outside the COMPASS range in $x_B$. The origin of this slope change remains to be understood, but it may simple be due to a statistical fluctuation.

\section{Summary}

In this talk, we have reviewed the HMCs correction scheme discussed in Refs.~\cite{Accardi:2009md,Guerrero:2015wha,Guerrero:2017yvf} and applied it to integrated Kaon multiplicities at HERMES and COMPASS beam energies. In this scheme, the mass effects are captured in a gauge invariant way by massive scaling variables which consider the need for the struck quark to have enough virtuality to fragment into a massive hadron. At LO in perturbation theory, and at LT, the finite-$Q^2$ cross section still factorizes as a product of PDFs and FFs, but evaluated at the Natchmann variable $\xi_h$ of Eq.~(\ref{eq:xi_h}) and the fragmentation scaling variable of Eq.~\eqref{eq:zeta}, respectively. 

After accounting for HMCs in this way, we found that the discrepancy in size between the measurements made by the HERMES and COMPASS collaborations is reduced. 
For the summed  $K^+ + K^-$ multiplicity there is still some difference in slope that
need to be investigated. In the case of the multiplicities $K^+/K^-$ ratio, the slopes of the two measurements become compatible. The last two $x_B$ bins for HERMES still show a suspicious behavior as it happen for the case of the sum $K^+ + K^-$, that could be partly attributed to nuclear binding and Fermi motion effect in the Deuteron target. However, nuclear effects should largely cancel out $K^+/K^-$ ratio, and the physical origin of the change in slope remains to be understood. It would be interesting to repeat this measurements at JLab 12, where a higher $x_B$ range can be covered at $Q^2$ values comparable to those probed at HERMES.

As an outlook, we would like to include nuclear corrections in our analysis to see if this explain the large $x_B$ behavior for HERMES, as well as prove factorization at NLO in perturbation theory for the case of a non vanishing virtuality for the fragmenting quark, $v\,'\,^2 \neq 0$. The results presented in this talk also points toward the necessity to use HMCs in fits of FFs \cite{deFlorian:2017lwf,Borsa:2017vwy,Ethier:2017zbq} including HERMES and COMPASS data.


\acknowledgments

We thank C.~Van Hulse, A.~Bressan, W.~Melnitchouk and N.~Sato for helpful discussions. This work was supported by the DOE contract No. DE-AC05-06OR23177, under which Jefferson Science Associates, LLC operates Jefferson Lab and DOE contract DE-SC0008791.

\end{document}